\documentclass[referee]{aa} 
\usepackage{graphicx}
\usepackage{natbib}
\usepackage{txfonts}
\usepackage{color}

\begin{document}
   \title{Comparison of the thin flux tube approximation with 3D MHD
simulations}


   \author{L. Yelles Chaouche\inst{1,2}
         \and
          S. K. Solanki\inst{1,3}
          \and
          M. Sch\"{u}ssler\inst{1}
          }


   \institute{Max-Planck-Institut f\"{u}r Sonnensystemforschung, 
Max-Planck-Strasse 2, 37191 Katlenburg-Lindau, Germany
   \and Instituto de Astrof\'{\i}sica de Canarias, C/ V\'{\i}a 
L\'{a}ctea,
   s/n, E38205 - La Laguna (Tenerife). Espa\~{n}a
   \and School of Space Research, Kyung Hee University, Yongin, 
Gyeonggi 446-701, Korea\\
              \email{yelles@mps.mpg.de}
             }



  \abstract
{The structure and dynamics of small vertical photospheric magnetic flux
concentrations has been often treated in the framework of an approximation
based upon a low-order truncation of the Taylor expansions of all quantities
in the horizontal direction, together with the assumption of instantaneous
total pressure balance at the boundary to the non-magnetic external medium.
Formally, such an approximation is justified if the diameter of the structure
(a flux tube or a flux sheet) is small compared to all other relevant length
scales (scale height, radius of curvature, wavelength, etc.). The advent of
realistic 3D radiative MHD simulations opens the possibility of checking the
consistency of the approximation with the properties of the flux
concentrations that form in the course of a simulation.} 
{We carry out a comparative analysis between the thin flux
   tube/sheet models and flux concentrations formed in a 3D
   radiation-MHD simulation.} 
{We compare the distribution of the vertical and horizontal components of the
magnetic field in a 3D MHD simulation with the field distribution in the case
of the thin flux tube/sheet approximation. We also consider the total (gas
plus magnetic) pressure in the MHD simulation box.} 
{Flux concentrations with super-equipartition fields are reasonably well
reproduced by the second-order thin flux tube/sheet approximation. The
differences between approximation and simulation are due to the asymmetry and
the dynamics of the simulated structures.}
{}

 \authorrunning{Yelles Chaouche et al.}
 \titlerunning{Thin flux tube approximation and 3D MHD simulations}
   \maketitle
%

\section{Introduction}

Much of the solar photospheric magnetic flux exists in the form of discrete
concentrations in intergranular lanes having a field strength of 1-2 kG
\citep{Stenflo:1973,Wiehr:1978, Rueedi:1992, Rabin:1992,Martinez:1997}, for
reviews see \citet{Solanki:1993}; \citet{Solanki:2006}.

Theoretical models of these flux concentrations have widely used the concept
of the flux tube: a bundle of field lines with circular cross-section
separated from the non-magnetic environment by a tangential discontinuity
\citep[see e.g.][]{Schuessler:1992}. Such a structure can be described, under
certain conditions, by the so called "thin flux tube approximation ". In its
simplest form, the axial component of the magnetic field is assumed to be
constant across the tube's cross-section, while the radial component is a
linear function of the radial coordinate \citep{Defouw:1976, Roberts:1978,
Roberts:1979}. The thin flux tube approximation can be formally justified if
the diameter of the flux tube is sufficiently small compared to variations of
the relevant physical quantities (such as pressure, density,$...$etc.) along
the tube's cross-section \citep{Spruit:1981, Schuessler:1992}.

The equations describing a thin flux tube can be obtained by writing all
physical quantities (magnetic field, temperature, pressure,$...$etc) in terms
of a Taylor expansion in the radial distance from the axis, and inserting
them in the MHD equations. By collecting terms of similar order one obtains a
hierarchy of equations \citep{Ferrizmas:1989a}. Truncating this hierarchy
after the $1^{st}$ order allows the $0^{th}$-order approximation introduced
above to be obtained.

Extensions of the thin flux tube approximation to higher orders have been
given in the literature. By retaining second-order terms \citet{Pneuman:1986}
have included in their modelling the effects of field line curvature,
internal structures, twist, and the merging of flux tubes with their
neighbours. A derivation of linear wave modes of a flux tube up to second
order has been carried out by \citet{Ferrizmas:1989}.

There is a large body of work in the literature based upon the thin flux tube
approximation. This includes theoretical work (structure of flux
concentrations, equilibrium, oscillations/wave, stability,...etc.) and
interpretation of observations \citep[for reviews, see][]{Solanki:1993,
Solanki:2006}. Various aspects of the thin tube approximation have been
compared to observational data
\citep[e.g.][]{Zayer:1989,Bruls:1995,Solanki:1996}, but its validity has not
been tested on the basis of the most advanced numerical simulations.

In the last two decades, the possibilities to self consistently model
magneto-convection at the solar photosphere using the full set of MHD
equations including radiative and convective energy transport
\citep[e.g.][]{Nordlund:1983, Stein:1998, Bercik:2002, Stein:2003,
Voegler:2003,Voegler:2005} have greatly improved. The structure of flux
concentrations in such MHD simulations appears rather complex, owing to their
interaction with convection and energy exchange with the neighbouring plasma.

We aim to evaluate to which extent the magnetic structures forming in 3D MHD
simulations can be described using the thin flux tube/sheet approximation.


\section{A series expansion of the thin flux tube/sheet equations} \label{sec:2}

We consider a magnetic flux tube to be a bundle of magnetic field lines with
a circular cross section, which is separated from its non-magnetic
surroundings by a tangential discontinuity with a surface current. For an
axisymmetric vertical flux tube, we adopt cylindrical coordinates
($r$,$\theta$,$z$), with the $z$-axis pointing in the vertical direction.
Physical quantities are regular at the axis ($r=0$), so that they can be
described in terms of a Taylor expansion in the radial coordinate
\citep{Roberts:1978, Spruit:1981, Pneuman:1986, Ferrizmas:1989,
Ferrizmas:1989a}

The properties of the axisymmetric MHD equations \citep{Ferrizmas:1989a}
imply that only even orders are non-zero in the above-mensioned expansions
for scalar quantities (such as temperature or density) and for $z$-components
of vectors, whereas for the radial and $\theta$-components of vectors only
the odd orders remain.

The three components of the magnetic field vector, the temperature and the
pressure can be written in a non-dimensional way:

\begin{equation}
b_{z}= h_{0}+ h_{2} x^{2} + h_{4} x^{4} + ...,\label{eqn:ser1}
\end{equation}

\begin{equation}
b_{r}= f_{1} x + f_{3} x^{3} + f_{5} x^{5} + ...,\label{eqn:ser2}
\end{equation}

\begin{equation}
b_{\theta}= s_{1} x + s_{3} x^{3} + s_{5} x^{5} + ...
\end{equation}

\begin{equation}
p= p_{0}+ p_{2} x^{2} + p_{4} x^{4} + ...
\end{equation}

\begin{equation}
\sigma= \sigma_{0}+ \sigma_{2} x^{2} + \sigma_{4} x^{4} +
...,\label{eqn:ser5}
\end{equation}
with $p=P/P^{*}$, $\sigma=T/T^{*}$, $b=B/B^{*}$, $x=r/H^{*}$, $y=z/H^{*}$,
$H^{*}= k T^{*}/(m_{p} g)$ and $\alpha=4 \pi P^{*}/B^{* 2}$. Where $B_{r}$,
$B_{\theta}$, $B_{z}$, represent the three components of the magnetic field
vector. $P$ and $T$ are the gas pressure and temperature, respectively. The
quantities with an asterisk are defined at the tube's axis ($x=r=0$) and at a
reference height ($z=y=0$). $k$ is Boltzmann's constant, $m_{p}$ the mean
particle mass, $g$ the gravitational acceleration, and $H$ the scale height.

\subsection{$B_{z}$ and $B_{r}$ under the thin flux tube approximation}\label{subsec:1}

Following, e.g., \citet{Pneuman:1986}, in a static atmosphere, we insert the
expansions~(\ref{eqn:ser1} to~\ref{eqn:ser5}) in the three components of the
momentum equation and the solenoidality relation, and collect terms of equal
power in $x$ into equations of corresponding order. Considering equations
including terms up to the third order, and assuming that the flux tubes
studied here have negligible twist, we obtain the following relations
\citep{Pneuman:1986}:

\begin{equation}
h_{2}=-\frac{1}{4} h''_{0}-\frac{\alpha p_{2}}{h_{0}},\label{eqn:x31}
\end{equation}

\begin{equation}
f_{1}=-\frac{1}{2} h'_{0},\label{eqn:x32}
\end{equation}
and

\begin{equation}
f_{3}=-\frac{1}{4} h'_{2},\label{eqn:x33}
\end{equation}
where the prime indicates a derivative with respect to $y$

We can then deduce $B_{z}$ up to the second order, and $B_{r}$ up to the
third order.

In order to close the above system it is necessary to consider relations
expressing magnetic flux conservation through the tube's cross-section and
total pressure balance at the boundary of the flux tube at any height
\citep{Ferrizmas:1989a, Ferrizmas:1989}. In addition to these relations,
\cite{Ferrizmas:1989} have considered an energy equation, whereas
\cite{Pneuman:1986} have chosen to prescribe two quantities, such as
$\sigma_0$ and $\sigma_2$, which allows more flexibility in defining the
atmosphere. In order to construct a thin flux tube which we will compare with
flux concentrations in MHD simulations, we take $h_0$ and $p_2$ from the MHD
simulations. Then $B_z$ and $B_r$ are determined from
equations~(\ref{eqn:x31}) to~(\ref{eqn:x33}). The cross section of the flux
tube is determined through the magnetic flux conservation relation:

\begin{equation}
\int_{0}^{x_0} (h_{0} + h_{2} x^{2}) dx =
\mathrm{Flux~at~base~of~tube} = \mathrm{constant},
\end{equation}
where $x_0$ is the tube's radius at a given height. The total pressure
balance can be expressed as:

\begin{equation}
(\beta p_i + b_{i}^{2})|_{x=x_0}=(8 \pi P_e +
B_{e}^{2})/B^{*2}|_{x=x_0},\label{eqn:ptot}
\end{equation}
where $\beta=8 \pi P^{*}/B^{* 2}$, the suffixes $i$ and $e$ indicate internal
and external quantities respectively and capital letters indicate dimensional
quantities.

Under the $0^{th}$-order approximation Eq.~(\ref{eqn:ptot}) reduces to:

\begin{equation}
(\beta p_0 + h_{0}^{2})|_{x=x_0}=(8 \pi P_e +
B_{e}^{2})/B^{*2}|_{x=x_0},\label{eqn:ptot0}
\end{equation}
This relation does not depend on the radius of the flux tube. Thus the total
pressure at a given height under the $0^{th}$-order approximation is constant
across the tube's cross-section.

Under the $2^{nd}$-order approximation we get:

\begin{equation}
(\beta p_0 + h_{0}^{2}) + x^{2} (\beta p_2 + f_{1}^{2} + 2 h_0
h_2)|_{x=x_0}=(8 \pi P_e +
B_{e}^{2})/B^{*2}|_{x=x_0},\label{eqn:ptot2}
\end{equation}
In this case the total pressure varies inside the flux tube, but has to match
the external total pressure at the tube's boundary.

\subsection{$B_{z}$ and $B_{x}$ under the thin flux sheet approximation}\label{subsec:2}

A flux sheet is an elongated structure with a small width (which we refer to
as $"W"$) compared to its length ($"L"$) along the solar surface, i.e. $W <<
L$ at the solar surface. A similar approach as described in the previous
section can be used to describe a thin flux sheet. In this case the magnetic
field component parallel to $L$ is constant, and thus plays no direct role in
the hydrostatic equilibrium. We can then adopt a Cartesian $2D$ geometry in
the $x-z$ plane, where $z$ is the vertical coordinate and $x$ is the
horizontal coordinate perpendicular to the vector $L$.

In a similar way to Sect.~\ref{subsec:1}, we can determine $h_{2}$, $f_{1}$
and $f_{3}$ as functions of $h_{0}$ and $p_{2}$.

\begin{equation}
h_{2}=-\frac{1}{2} h''_{0}-\frac{\alpha p_{2}}{h_{0}}
\end{equation}

\begin{equation}
f_{1}=- h'_{0}
\end{equation}
and

\begin{equation}
f_{3}=-\frac{1}{3} h'_{2}
\end{equation}

Note the similarity between these equations and the ones describing the thin
flux tube. The main difference (apart of the geometry) is the numerical
values of the constant coefficients which affects, for instance, the
expansion rate of the flux tube/sheet with height.

\section{The radiative MHD simulations} \label{sec:3}

Three dimensional radiation-MHD simulations of the solar photosphere have
been described by \citep{Nordlund:1983, Nordlund:1990, Stein:1998,
Bercik:2002, Stein:2003, Voegler:2003, Voegler:2005}.

\begin{figure}
\includegraphics[width=0.95\linewidth,bb= 80 125 520 600]{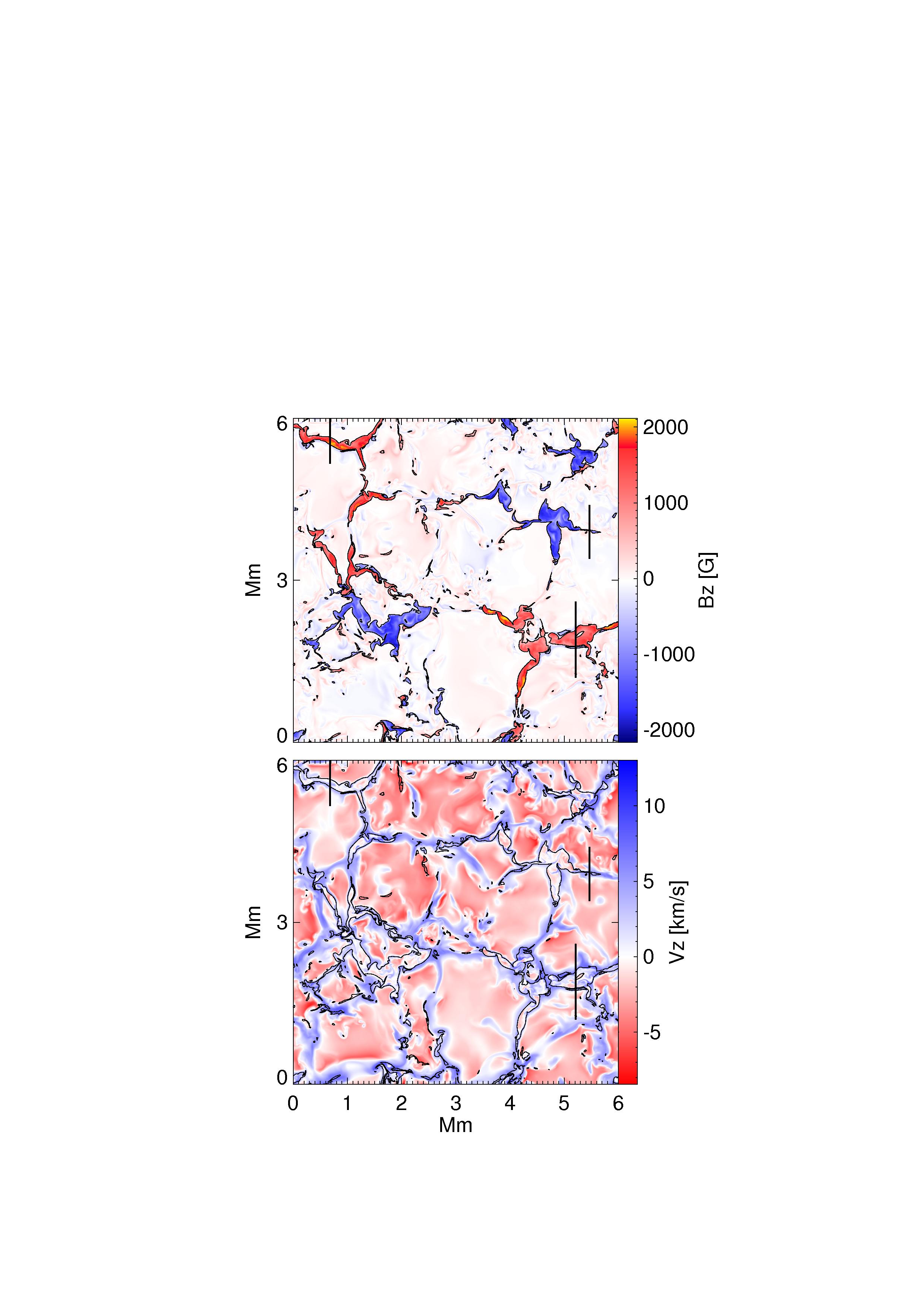}
\caption{Upper panel: vertical component of the magnetic field ($B_{z}$) at a
fixed geometrical height near the averaged visible solar surface
($\tau_{5000}=1$). Lower panel: vertical component of the velocity ($V_{z}$)
at $\tau_{5000}=1$. Downflows are represented in red and upflows in blue. The
black contours outline regions where $|B_{z}| > 500$ G. The black vertical
lines indicate locations where we carry out a detailed analysis of magnetic
elements in Section~\ref{sec:44}.\label{fig:9b}}
\end{figure}

The simulation run used here has been obtained with the fully compressible
MURaM code \citep{Voegler:2003thesis, Voegler:2005}. It takes into account
non-local and non-gray radiative energy transport, and includes the effects
of partial ionization. The simulation box has a horizontal extension of
$6\times6$ $\mathrm{Mm}^{2}$ and is $1.4$ Mm deep. The grid cell size is 5 km
in the horizontal direction and 7 km in the vertical. The simulation run
starts from a plane-parallel atmosphere which extends from $-0.8$ Mm below to
$0.6$ Mm above a reference $0$, which is roughly situated $-100$ km below the
average continuum optical depth unity ($\tau_{5000}=1$, which corresponds to
the solar surface at $5000$ {\AA}). After convection has fully developed, a
mixed-polarity magnetic field configuration with zero net vertical flux is
introduced. This is done such that the simulation domain is divided into four
parts with vertical field of alternate polarities in a chessboard pattern. We
choose a representative snapshot for our analysis (see Fig.~\ref{fig:9b}).
The mean unsigned field strength at optical depth unity is $200$ G for this
snapshot.

\section{Analysis of the total pressure in the whole simulation
domain} \label{sec:5}

\begin{figure}
\includegraphics[width=0.95\linewidth,bb= 105 158 510 480]{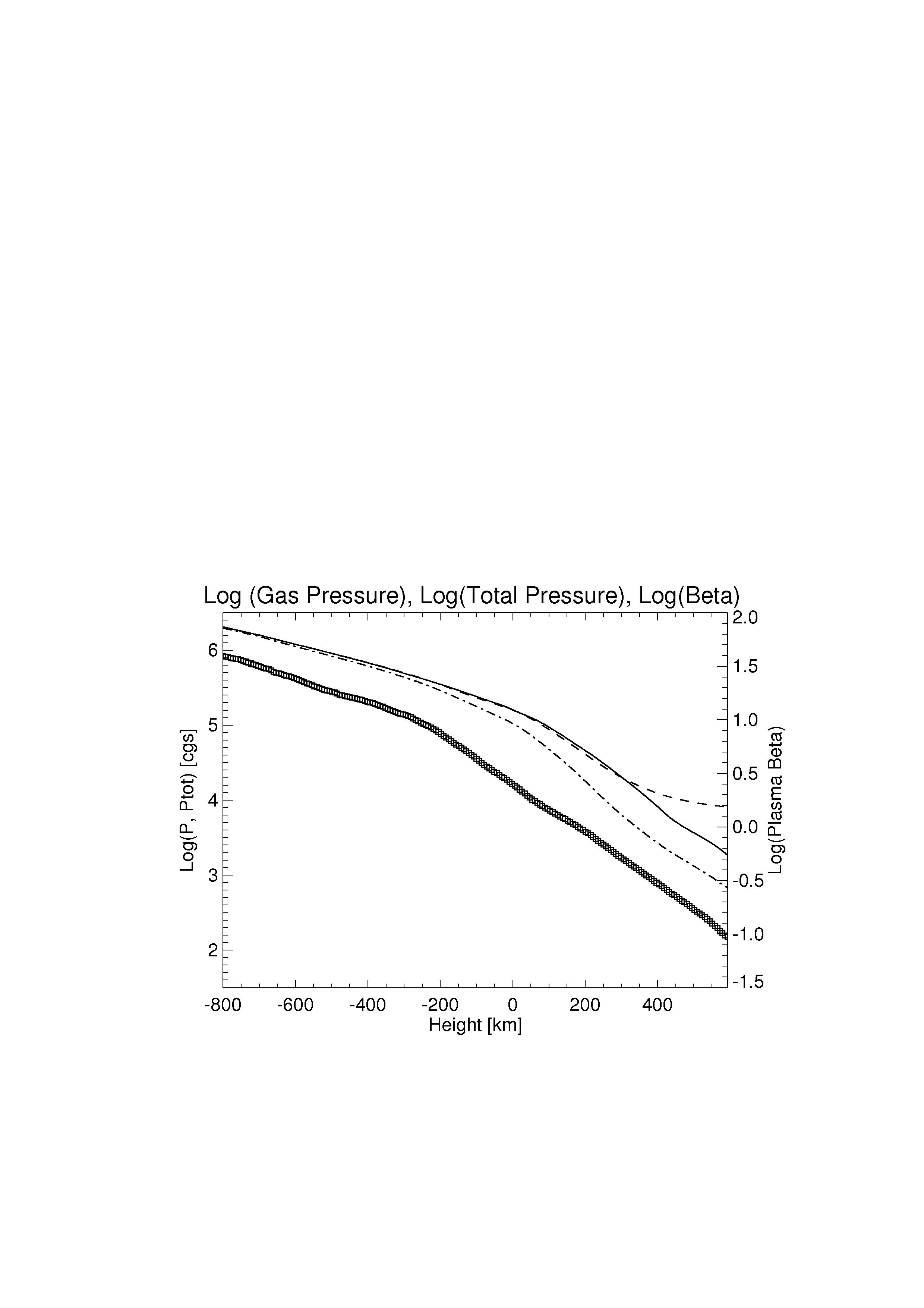}
\caption{Horizontal average of the plasma $\beta$, gas pressure and total pressure as a function of
height. Plotted are, the logarithmic values of the gas pressure averaged over
regions with field strength smaller than 50 G (full line), the gas pressure averaged in magnetic flux concentrations (dash-dotted line),
the total pressure over magnetic flux
concentrations (dashed line) and the plasma $\beta$ (squares; scale on the right). \label{fig:7}}
\end{figure}

Figure~\ref{fig:7} shows horizontally averaged gas and total pressures as a
function of height. The solid line represents the gas pressure averaged over
regions with field strength smaller than 50 G. The dash-dotted line indicates
the gas pressure averaged over magnetic regions. The threshold in $|B_{z}|$
defining magnetic regions varies linearly from $500$ G at the bottom of the
simulation box to $300$ G at the top. The dashed line represents the total
pressure ($P_{tot} $=$ P + B^{2}/(8 \pi)$) averaged over magnetic regions.
The plasma $\beta=8 \pi P/B^{2}$ for magnetic regions is indicated by square
symbols.

The difference between the gas pressures inside and outside magnetic regions
becomes smaller with depth. This is due to the large values of the plasma
$\beta$ in the deep layers (e.g. below $-400$ km) which indicate that the
pressure balance between magnetic features and their surroundings is mainly
ensured by gas pressure. Above $300$ km, the total pressure in magnetic
features shows an excess compared to gas pressure in nearly field-free areas.
This excess increases with height and is due to the effect of curvature
forces. This implies that the $0^{th}$-order thin flux tube/sheet
approximation is not sufficient to describe the flux concentrations in the
upper part of the simulation box. The gas pressure in nearly field-free
regions is higher than the total pressure in the flux concentrations in the
height-range situated between $50$ and $300$ km. This slight pressure
excess mainly results from the fact that the gas pressure at equal
geometrical height is, on average, higher in the granular upflows than in the
intergranular downflow lanes, where the magnetic flux concentrations reside.
In addition, a pressure deficit in the flux concentrations relative to their
local environment could arise as the result of the outward curvature force of
the expanding tubes between $0$ and $300$ km height. Above $300$ km, the sign
of the curvature force is reversed as a result of the wineglass shape of the
flux tubes caused by the presence of neighboring tubes (reflected in our
simulation by the vertical-field upper boundary condition). In any case, the
deviation from total pressure balance is very small below $300$ km height.

\begin{figure*}
\includegraphics[width=0.95\linewidth,bb= 55 105 520 550]{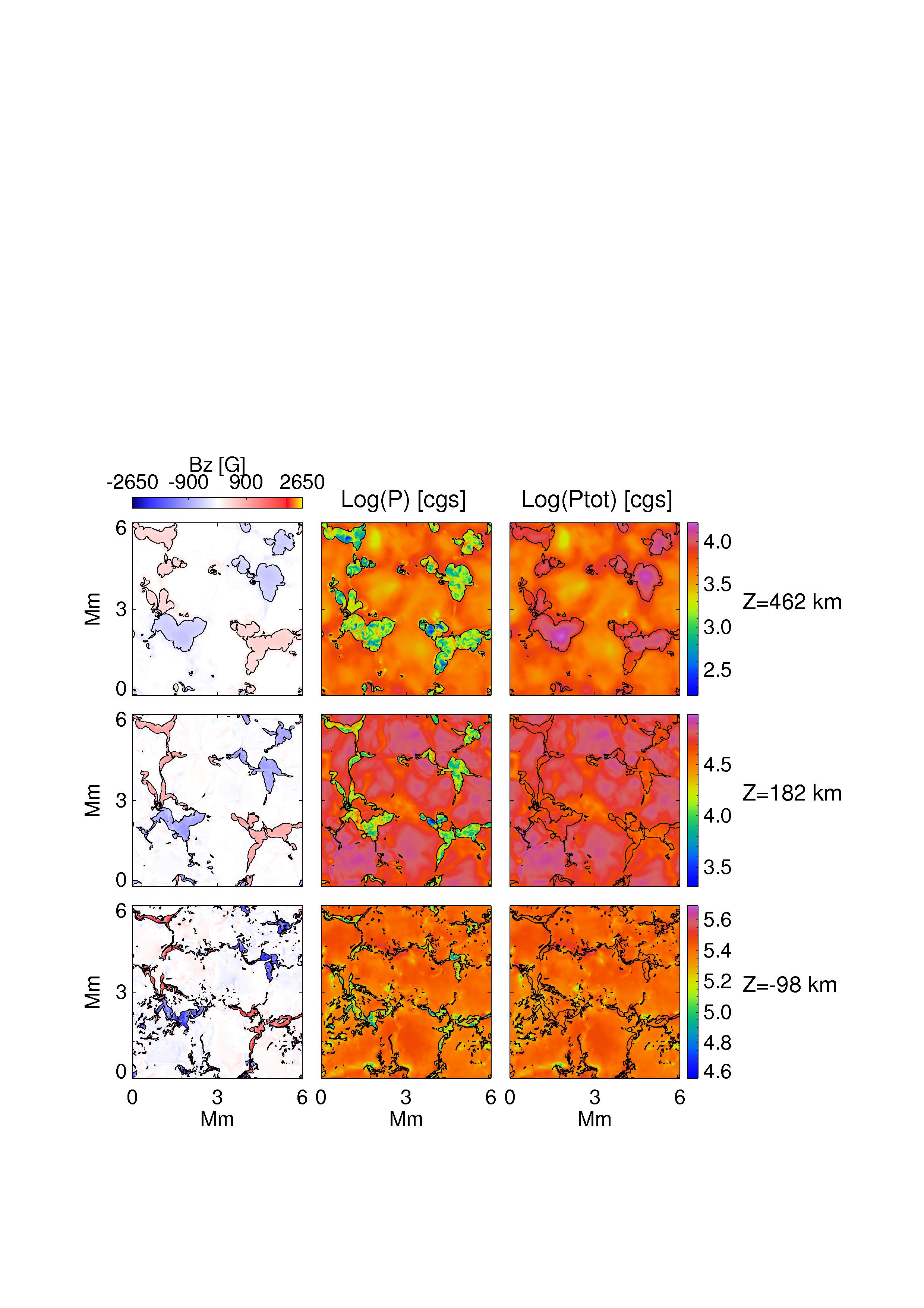}
\caption{Gas pressure, total pressure and vertical component of the magnetic
field at three geometrical heights $-98$, $182$ and $462$ km. The reference
height "$0$" is located at $-100$ km below the average continuum optical
depth unity at $5000$ {\AA}. A common color table is used for the three
$B_{z}$ maps. At each height, the pressures share the same color table
indicated on the right-side of the maps. The black contours enclose regions
where $|B_{z}|$ is higher than $500$ G at $-98$ km, higher than $400$ G at
$182$ km and higher than $300$ G at $462$ km.\label{fig:12}}
\end{figure*}

The total pressure balance between a magnetic flux concentration and its
non-magnetic surroundings results from the continuity of the normal stress at
the boundary separating the flux concentration from its surroundings. In the
$0^{th}$-order approximation (Eq~\ref{eqn:ptot0}), $P_{tot}$ not only matches
the boundary value but is also constant across the flux concentration. The
presence of $2^{nd}$-order terms (or higher-orders) produces higher or lower
values of $P_{tot}$ at the center of flux concentrations (Eq~\ref{eqn:ptot2})
in comparison to $P_{tot}$ at the magnenic/non-magnetic boundary, which
remains equal to the external pressure. Thus the total pressure can be used
as diagnostic for determining whether a flux concentration has $0^{th}$ or
higher-order configuration. In order to illustrate the distributions of
pressures and magnetic field in the simulation box, which includes different
sizes and shapes of magnetic concentrations, we show in Figure~\ref{fig:12}
maps of the gas and total pressures as well as $B_{z}$ at three heights
$-98$, $182$ and $462$ km, where the reference height $0$ is roughly situated
$-100$ km below the average continuum optical depth unity, $\tau_{5000}=1$.

The vertical component of the magnetic field at $-98$km is displayed in the
lower left pannel of Figure~\ref{fig:12}. Note that the flux resides mainly
in strong flux concentrations located in intergranular lanes. The middle
panel of the lower row in Figure~\ref{fig:12} represents gas pressure at
$-98$ km. Locations where the magnetic flux density is higher than 500 G are
outlined by dark contours. The gas pressure is higher at centers of granules
comparing to intergranules. This pressure excess drives the horizontal flows
towards the intergranular lanes \citep[see e.g.][]{Stein:2003}. Intergranular
lanes display a mixed picture with high gas pressure (which stops the
horizontal flows) but also low pressure areas. The magnetic flux
concentrations show lower gas pressure.

The total pressure inside flux concentrations at $-98$ km (lower right panel
of Figure~\ref{fig:12}) is roughly close to gas pressure outside, and does
not vary significantly within individual flux concentrations. Constant
$P_{tot}$ is a necessary condition (but not sufficient) for the validity of
the $0^{th}$-order thin flux tube/sheet model.


The existence of $2^{nd}$-order terms (or higher-orders) in pressure and
magnetic field leads to higher/lower values of the total pressure at the
center of flux concentrations. So, one way of measuring the importance of
higher-order terms is to compute the standard deviation and the mean value of
the total pressure inside magnetic elements and compare them with the
corresponding values outside magnetic regions (see Table~\ref{table:2}).

\begin{table*}[!ht]
\caption{Standard deviations and mean value of $P_{tot}$}

\begin{center}
\normalsize
\begin{tabular}{l c c c c c c}

 \hline
Altitude [km] & -98 & 182 & 462 \\
Standard deviation of $P_{tot}$ in non-magnetic regions  ($\sigma_{NM}$) [cgs]& 27703.0  &  9707.59  & 1055.04 \\
Standard deviation of $P_{tot}$ in magnetic regions  ($\sigma_{M}$) [cgs]& 32709.4  & 7950.80  & 2151.41 \\
Mean value of $P_{tot}$ in non-magnetic regions ($\overline{NM}$) [cgs]& 254220.0 &  57750.9  & 4901.33 \\
Mean value of $P_{tot}$ in magnetic regions ($\overline{M}$) [cgs]& 249696.0 & 47409.8  &  8626.20 \\
$\sigma_{NM}$/$\overline{NM}$ & 0.108 & 0.168  & 0.215 \\
$\sigma_{M}$/$\overline{M}$ & 0.130 & 0.167  & 0.249 \\
 \hline
 \end{tabular}
 \label{table:2}
 \end{center}
 \end{table*}

At an altitude of $-98$ km (Table~\ref{table:2}), $\overline{M}$ is slightly
lower than $\overline{NM}$ because magnetic flux concentrations are located
in intergranular lanes where the pressure at this altitude is slightly lower
than the average pressure over the simulation domain.
$\sigma_{M}$/$\overline{M}$ is larger than $\sigma_{NM}$/$\overline{NM}$,
this does not result from higher-order terms, but rather indicates the
presence of fluctuations inside magnetic elements. This is due to the fact
that the plasma beta at this altitude is larger than unity (See
Fig.~\ref{fig:7}) which indicates that convection affects and perturbs the
field's regularity. Note that locations with particularly low total pressure
(e.g. the green-colored ones) are generally unrelated to magnetic flux
concentrations.

At a higher altitude ($182$ km) we see in Figure~\ref{fig:12} that magnetic
structures have expanded. The gas pressure has, on average, lower values in
intergranular lanes and particularly low values inside magnetic elements. The
total pressure is lower in intergranular lanes even when there is no (or low)
magnetic field, e.g. in the region around the coordinates (3Mm, 2.8Mm). The
mean value $\overline{NM}$ is higher than $\overline{M}$
(Table~\ref{table:2}). The normalized fluctuations of $P_{tot}$ inside and
outside magnetic elements are similar
($\sigma_{M}$/$\overline{M}$$\approx$$\sigma_{NM}$/$\overline{NM}$). Hence
there is little evidence for a significant contribution from higher-order
terms.


Near the top of the box, at a geometrical height of $462$ km, we notice that
the total pressure (Figure~\ref{fig:12}) increases towards the center of flux
concentrations. $\overline{M}$ $>$ $\overline{NM}$ and
$\sigma_{M}$/$\overline{M}$ $>$ $\sigma_{NM}$/$\overline{NM}$. This indicates
that the total pressure is not a $0^{th}$-order function. This effect is more
pronounced in large flux concentrations. The plasma $\beta$ is small at these
heights (see Fig.~\ref{fig:7}), thus we expect a nearly force-free
equilibrium with a balance between curvature force and magnetic pressure
gradient. So the outward magnetic pressure force will be balanced by the
inward curvature force. Thus the magnetic pressure ($\simeq$ total pressure)
has to increase inward. Hence the increase of $P_{tot}$ at the center of flux
concentrations in the upper right panel of Figure~\ref{fig:12}.

\section{Analysis of individual magnetic structures} \label{sec:44}

The flat profiles of total pressure in the lower part of the atmosphere are
in favour of the applicability of the $0^{th}$-order thin flux tube/sheet
approximation. In the upper part of the atmosphere, however, the magnetic
features show a total pressure excess in their center. This indicates that
the $0^{th}$-order thin flux tube/sheet approximation is not applicable, but
possibly the extension of the approximation to $2^{nd}$-order is sufficient
to describe the force equilibrium of the magnetic structures. For a
quantitative investigation we select three flux concentrations in the MHD
simulation run according to their width and morphology. These will be treated
in the next three sub-sections.

\subsection{Thin flux sheet} \label{sec:4}

We compare the properties of a narrow flux sheet in the MHD simulation
(Fig.~\ref{fig:9b}) with the thin flux sheet model presented in
Sect.~\ref{subsec:2}. Note that the flux tubes/sheets in a magneto-convection
simulation are not static (unlike the assumption made in Sect.~\ref{sec:2}).
They interact with the external plasma, and get distorted by the granulation
motion. They also exchange energy (mainly by radiation) with the
surroundings. In order to maintain the numerical stability of the simulation,
the gradient of any physical quantity cannot be too large between two
neighbouring grid cells. More specifically, the magnetic flux density must
not jump abruptly from the boundary of a flux tube to the neighbouring
non-magnetized plasma \citep[][]{Voegler:2003thesis}. Thus the boundary layer
separating a flux tube from the surrounding non-magnetized plasma is a few
grid points wide, unlike the tangential discontinuity in the case of an ideal
flux tube. We wish to see whether simulations and thin flux sheet/tube
approximation are consistent with each other in spite of the fact that MHD
flux tubes/sheets have finite boundary layers, internal and external dynamics
and deviate from an axi- or translationally symmetric configuration.

\begin{figure}
\includegraphics[width=0.95\linewidth,bb= 85 108 500 610]{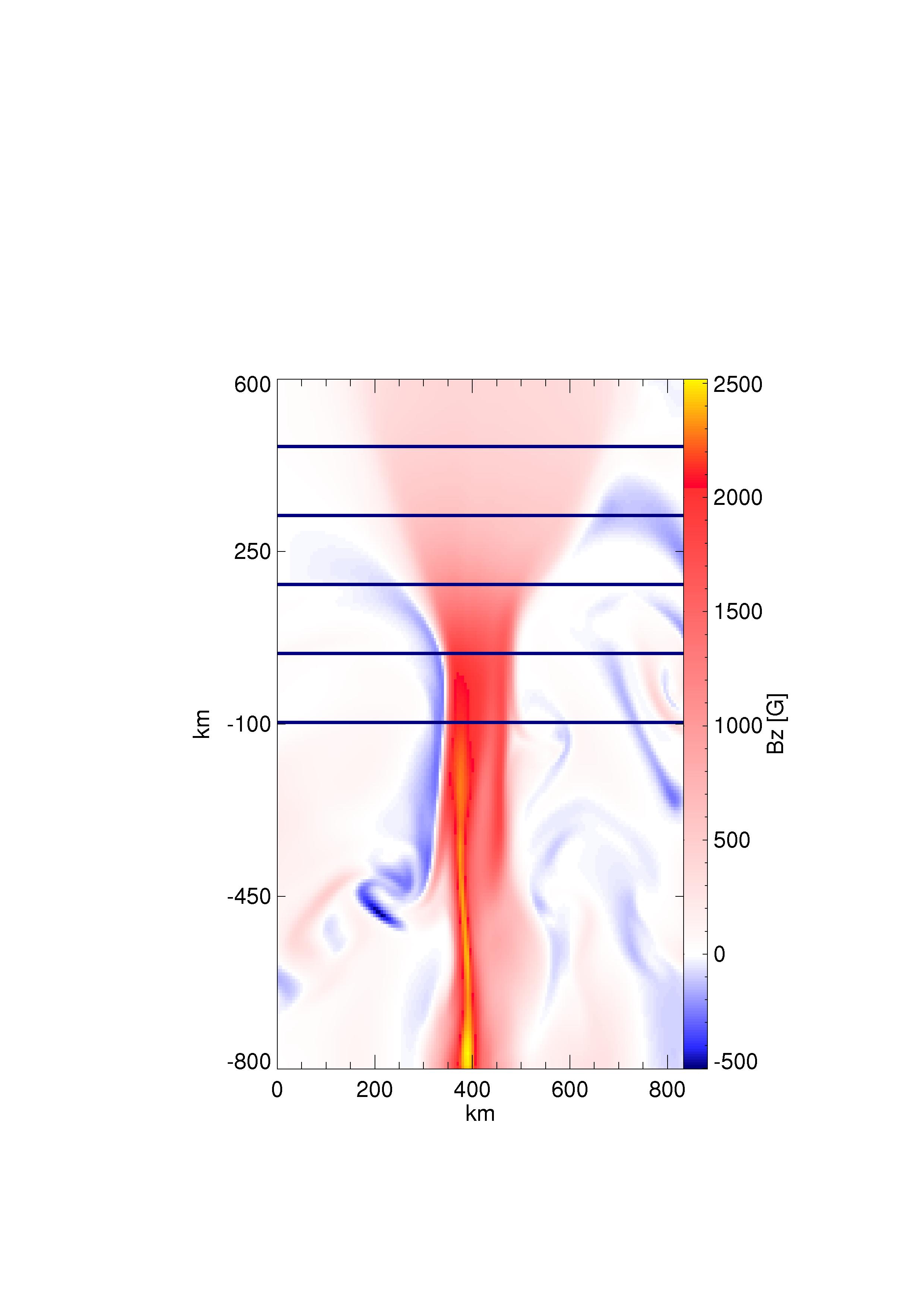}
\caption{Vertical 2D cut through the MHD simulation box at the location shown by
the black line at the upper-left corner in Figure~\ref{fig:9b}. The horizontal lines indicate
locations at which we plot profiles of various physical
quantities in Figs.~\ref{fig:6b} to~\ref{fig:5b}.\label{fig:2b}}
\end{figure}

We select a rather narrow sheet-like structure in the simulation domain. A
vertical 2D cut through the flux sheet (shown in Fig.~\ref{fig:2b}) at the
location indicated by the dark line in the upper left corner of
Fig.~\ref{fig:9b} reveals the morphology of the magnetic field. The expansion
of the flux sheet with height is mainly determined by magnetic flux
conservation with height and a horizontal balance between the magnetic plus
gas pressure inside the sheet with the gas pressure outside.

\begin{figure}
\includegraphics[width=0.95\linewidth,bb= 85 88 530 610]{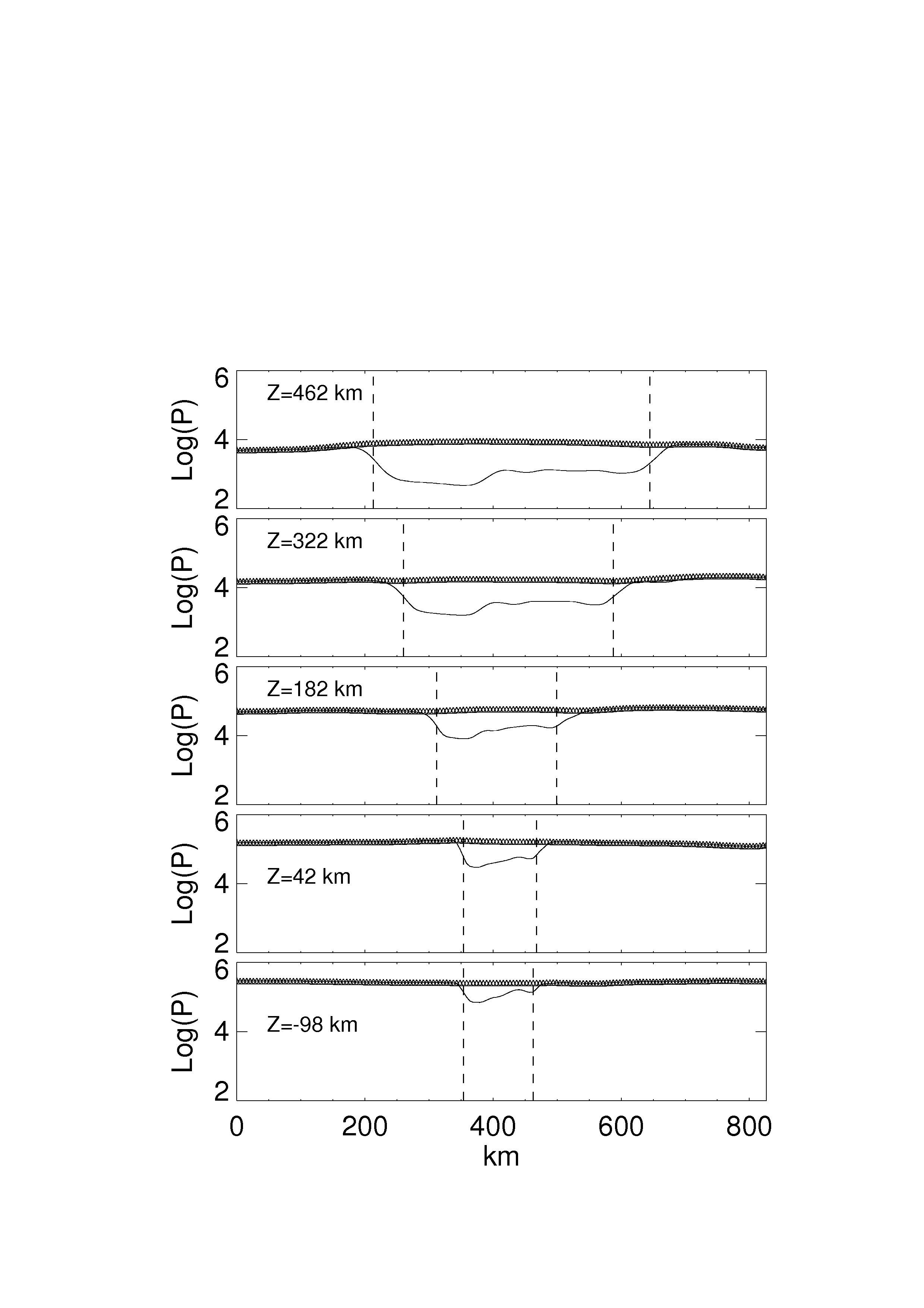}
\caption{Gas pressure (solid lines) and total pressure (triangles)
along the 5 horizontal lines in
Figure~\ref{fig:2b}. The vertical dashed lines outline regions where $B_{z}$ is
larger than $75$ percent of its maximum value.\label{fig:6b}}
\end{figure}

Figure~\ref{fig:6b} shows profiles of gas pressure (full lines) and total
pressure (triangles) along the $5$ horizontal lines in Figure~\ref{fig:2b}.
The location of the magnetic flux concentration is reflected by the lower gas
pressure. The vertical dashed lines outline regions where $B_{z}$ is higher
than $0.75$ of its maximum value. The profiles indicate that the flux sheet's
equilibrium in the lower panels is consistent with balance of total pressure
in the zeroth-order thin flux sheet approximation (see Eq~\ref{eqn:ptot0},
which is valid for both flux tubes and flux sheets). In the top panel we see
that the total pressure increases somewhat towards the center of the sheet,
which indicates the necessity of extending the approximation to second- (or
higher-) order (see Eq~\ref{eqn:ptot2}). At this height, the plasma $\beta$
has become so small that the internal equilibrium becomes nearly force free,
i.e., curvature forces and magnetic pressure gradient balance each other.

\begin{figure}
\includegraphics[width=0.95\linewidth,bb= 70 92 520 610]{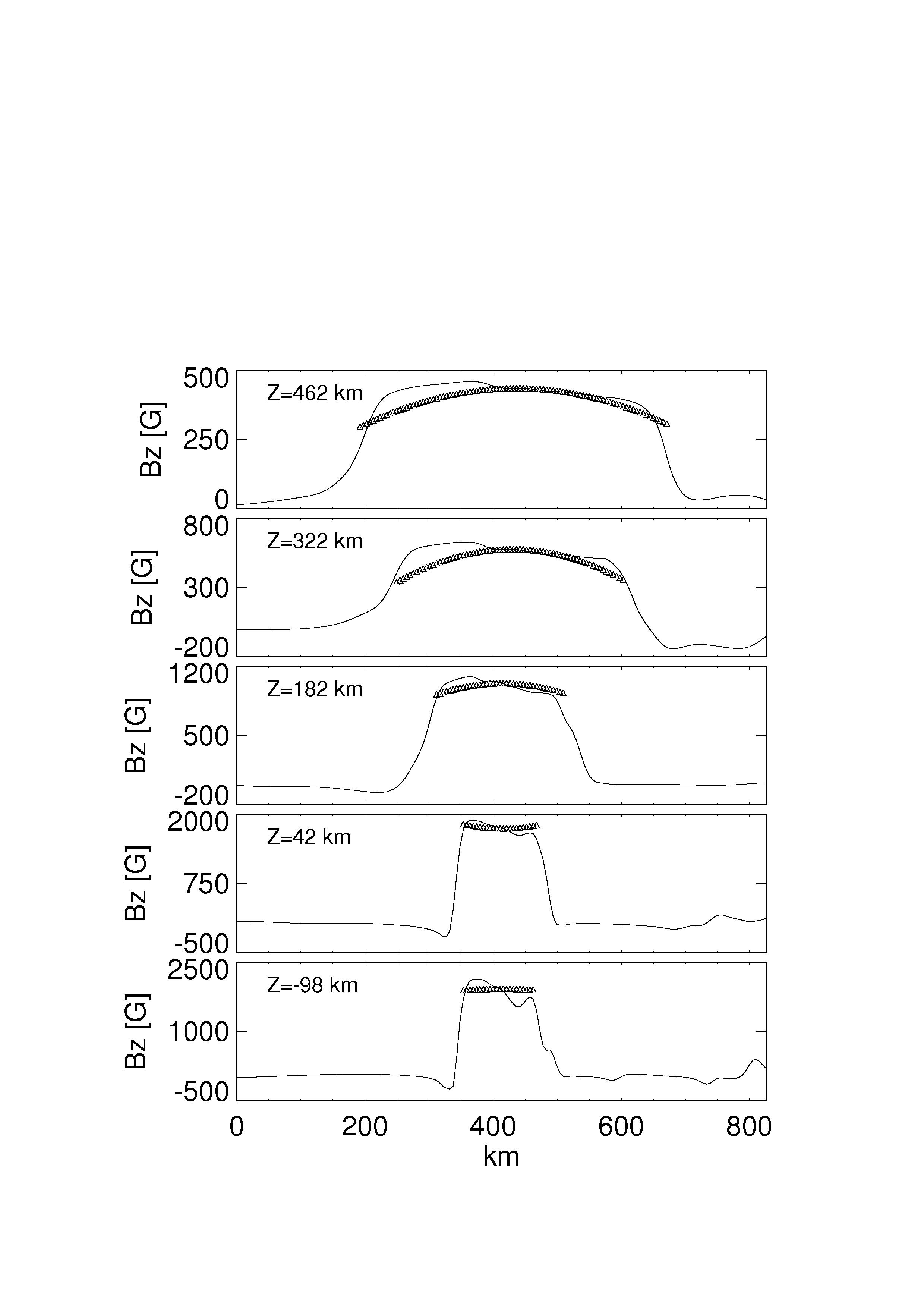}
\caption{Vertical component of the magnetic field, $B_{z}$, along the
$5$ horizontal lines in Figure ~\ref{fig:2b} (solid lines). The triangles
represent $B_z$ resulting from a $2^{nd}$-order thin flux sheet
model. \label{fig:4b}}
\end{figure}

Figure~\ref{fig:4b} shows the vertical component of the magnetic field along
the 5 horizontal lines in Figure~\ref{fig:2b}. The triangles represent
$B_{z}$ in the case of a thin flux sheet in the second-order approximation.
The solid curves represent $B_z$ from the MHD simulations. We note that
$B_{z}$ for the thin flux sheet at the two lower panels is close to a
constant (small contribution from the $2^{nd}$-order terms), whereas in the
three upper panels the second-order terms become more important.

The $2^{nd}$-order approximation reproduces reasonably well the overall
$B_{z}$ profiles obtained from the MHD simulations in the higher layers of
the atmosphere. The profiles of $B_z$ from the simulation exhibit some
structures across the sheet's cross-section which are not reproduced by the
thin sheet model. This is because this latter model produces only symmetric
profiles of $B_z$ \citep{Ferrizmas:1989a}. The actual profiles of $B_z$ are
asymmetric primarily in the sense that the left part exhibits larger values
than the right part. This is associated with lower values of the pressure at
these locations, so $B_z$ has to increase in order to keep $P_{tot}$ balanced
(see Figure~\ref{fig:6b}).

\begin{figure}
\includegraphics[width=0.95\linewidth,bb= 85 92 520 610]{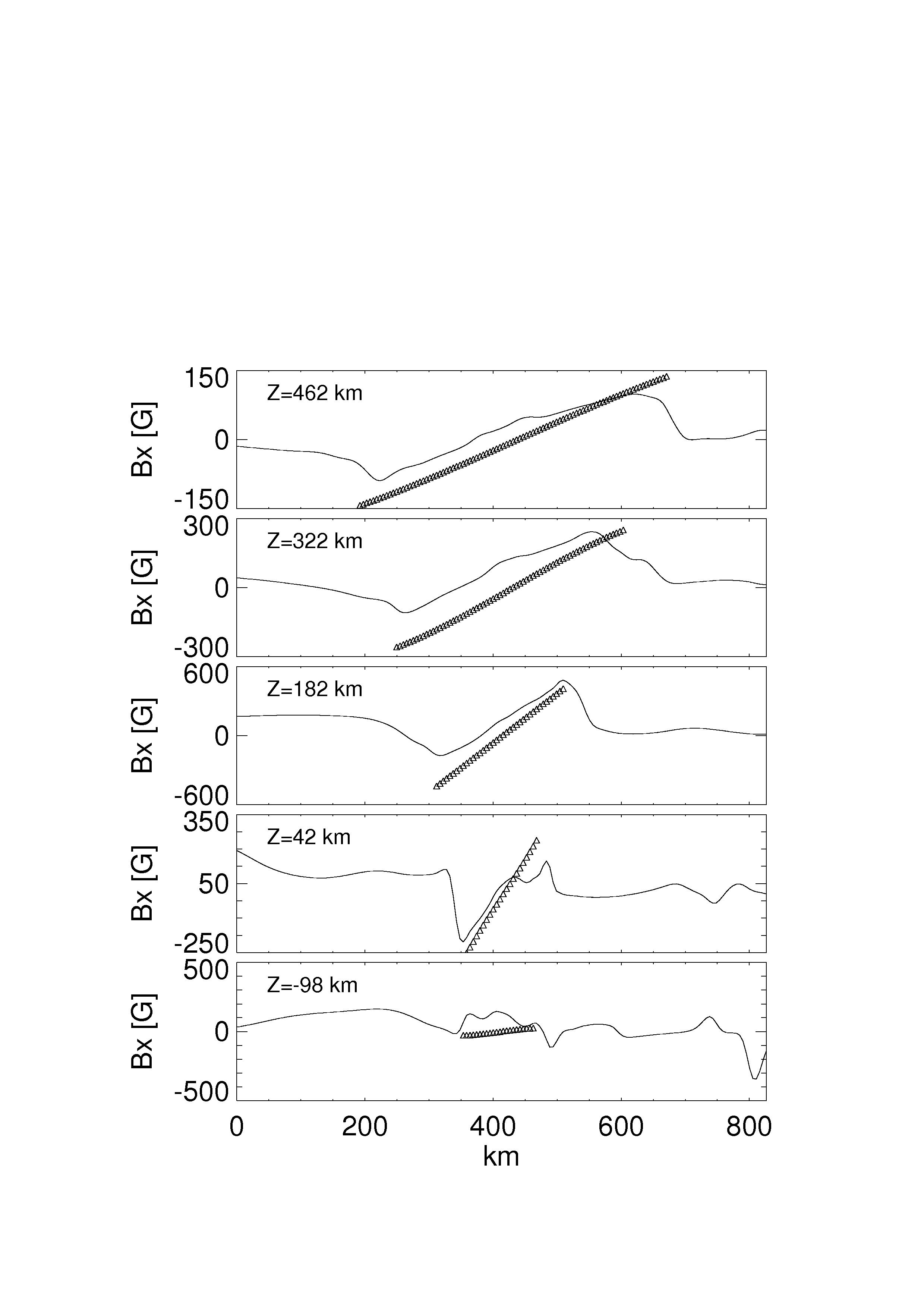}

\caption{Full lines: horizontal component of the magnetic field, $B_{x}$,
across the flux sheet plotted along the $5$ horizontal cuts in
Fig.~\ref{fig:2b}. The triangles represents $B_x$ resulting from a
$2^{nd}$-order thin flux sheet model.\label{fig:5b}}
\end{figure}

The distribution of the horizontal field component and its approximation with
the thin flux sheet model are shown in Fig.~\ref{fig:5b}. Here $B_x$ includes
a third-order term (see section~\ref{subsec:2}). The profiles of the actual
field are smooth for the three upper panels (low $\beta$). In the two lowest
panel we notice some fluctuations mainly due to perturbations by the external
convection. The fit between the simulation result and the thin sheet model is
relatively good for the three upper panels, and less good for the two lower
ones. We notice that for the three upper panels there is a systematic offset
between the actual values and the thin sheet model. This is due to the fact
that the sheet is slightly inclined towards the right (more positive $B_{x}$
than negative in Fig.~\ref{fig:5b}). This can also be seen in
Fig.~\ref{fig:2b} at heights above $\approx 150$ km.

$B_{z}$ and $B_{x}$ for the thin flux sheet can be written in a dimensional
way (see also Eqns~\ref{eqn:ser1} and~\ref{eqn:ser2}):

\begin{equation}
B_{z}= B_{z 0}+ x^{2} B_{z 2} + x^{4} B_{z 4} + ... = B0 + B2 + B4 +
... ,\label{eqn:serie_100}
\end{equation}
\begin{equation}
B_{x}=  x B_{x1} + x^{3} B_{x3} + ... = B1 + B3 +
...,\label{eqn:serie_200}
\end{equation}

We can compare the relative importance of successive terms in these series
expansions. Table~\ref{table:3} indicates that the average values
$\overline{|B2|}$ are significantly smaller than $\overline{|B0|}$. The
importance of $\overline{|B2|}$ is more pronounced in the upper part of the
atmosphere. This is also noticeable in the upper panels of
Figure~\ref{fig:4b}. In a similar way to Sect.~\ref{subsec:2} we calculate
$4^{th}$-order terms (see also \cite{Pneuman:1986} and
\cite{Ferrizmas:1989a}). Table~\ref{table:3} shows that $\overline{|B4|}$
terms are very small compared to $\overline{|B0|}$. Their relative importance
reaches its maximum in the top part of the atmosphere, though they remain
negligible in practical terms. Similarly, $\overline{|B3|}$ terms are very
much smaller than $\overline{|B1|}$. Thus the influence of successive terms
in Eqs.~(\ref{eqn:serie_100}) and~(\ref{eqn:serie_200}) decrease with their
order. This is clearly seen in Figures~\ref{fig:4b} and~\ref{fig:5b}, and
confirms that neglecting the $4^{th}$-order terms in $B_z$ is justified.

\begin{table*}
\caption{Relative importance of the average values of the series expansion
$\overline{|B2|}$/$\overline{|B0|}$, $\overline{|B4|}$/$\overline{|B0|}$ and
$\overline{|B3|}$/$\overline{|B1|}^{\mathrm{*}}$. } \label{table:3}

\begin{center}  
\begin{tabular}{c c c c c c}

 \hline
Thin flux sheet: height from reference [km] & -98 & 42 & 182 & 322 & 462 \\
$\overline{|B2|}$/$\overline{|B0|}$ & 0.009 & 0.014 & 0.036 & 0.118 & 0.106 \\
$\overline{|B4|}$/$\overline{|B0|}$ & 7.67 e-05 & 2.75 e-05 & 0.001 & 0.010 & 0.015 \\
$\overline{|B3|}$/$\overline{|B1|}$ & 0.086 & 0.028 & 0.006 & 0.056 & 0.035 \\
 \hline
Thick flux tube: height from reference [km] & -98 & 42 & 182 & 322 & 462 \\
$\overline{|B2|}$/$\overline{|B0|}$ & 0.112 & 0.149 & 0.007 & 0.086 & 0.059 \\
$\overline{|B4|}$/$\overline{|B0|}$ & 0.017  & 0.012  &  4.75 e-04 & 0.011 & 0.003 \\
$\overline{|B3|}$/$\overline{|B1|}$ & 0.183 & 0.142 & 0.094 & 0.023 & 0.034 \\
 \hline
 \end{tabular}

 \end{center}
\begin{list}{}{}
\item[$^{\mathrm{*}}$] Vertical bars indicate absolute values and overlines indicate horizontal
average over the sheet's cross-section
\end{list}
 \end{table*}


\subsection{Analysis of a broad flux concentration} \label{sec:6}

In this section, we compare $B_z$ and $B_r$ from a thick flux concentration
with the thin tube model (Sect.~\ref{subsec:1}). The criteria for the choice
of a flux tube in the MHD simulations are primarily its width and a relative
smoothness of $P_{tot}$ across it. The selected flux tube is located in the
lower right part of Fig.~\ref{fig:9b} (crossed by a dark line). The first
thing to note is that the tube is split near the solar surface, which
probably results from the history of its interaction with convection. We also
notice that this "tube" has a cross-section which deviates significantly from
a circular area (see Fig.~\ref{fig:9b}).

\begin{figure}
\includegraphics[width=0.95\linewidth,bb= 70 92 520 610]{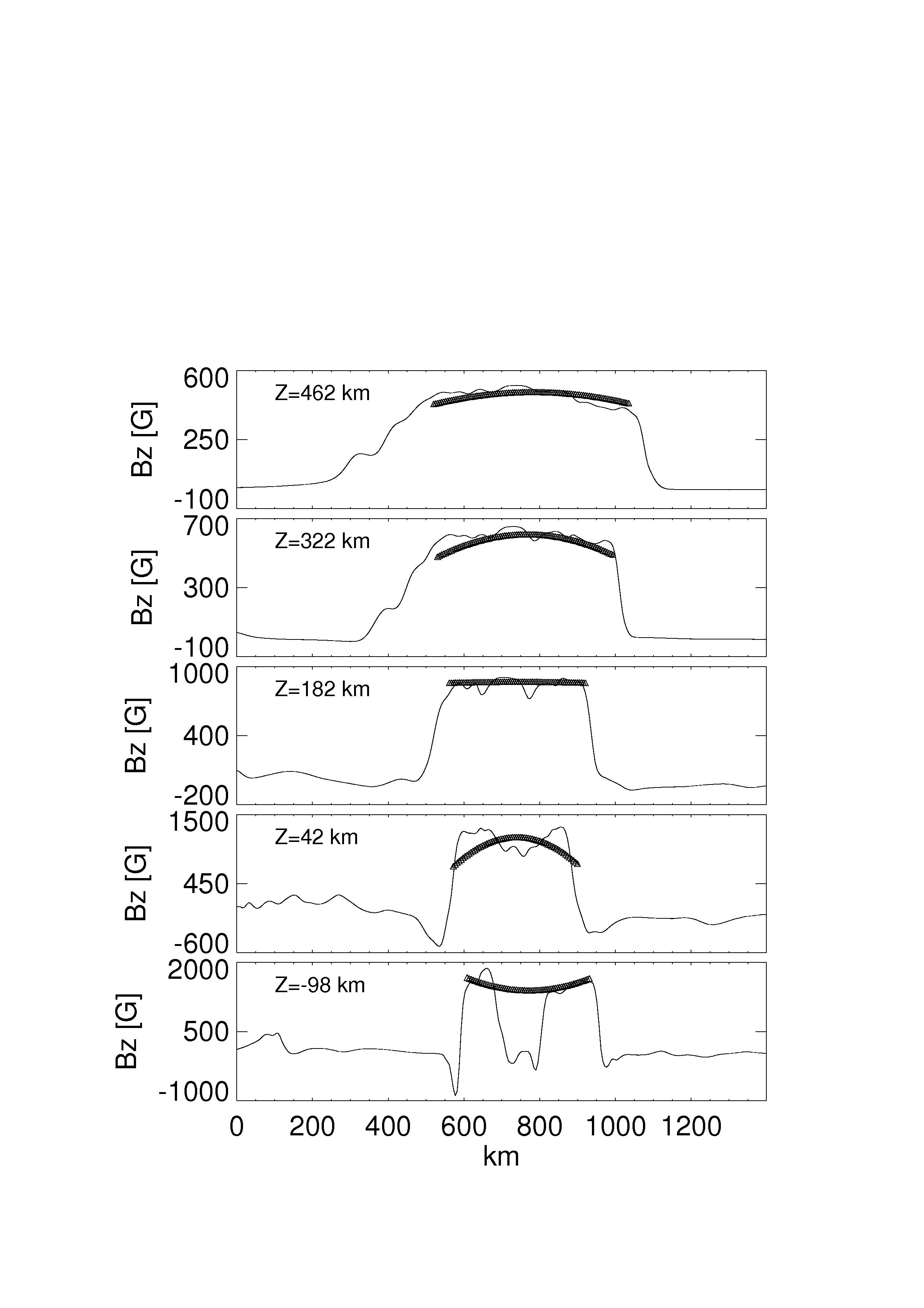}
\caption{Vertical component of the magnetic field of a thick flux tube in the MHD snapshot
(solid lines) along the black line in the lower right corner of Fig.~\ref{fig:12}.
The triangles represent $B_{z}$ resulting from the thin flux tube model.
The 5 plots correspond
to the same heights as in Fig.~\ref{fig:4b}.\label{fig:23b}}
\end{figure}

Inspite of these facts, the thin flux tube model reproduces reasonably well
the overall shape of $B_z$ given in the three upper panels of
Fig.~\ref{fig:23b}. In the upper two, we notice the existence of a region
with smoother decrease of $B_z$ at the left edge of the flux tube. This
results from a small neighbouring magnetic structure that merges with the
main flux tube. This structure is not visible in the lower panels since at
those heights it does not overlap the dark line (Fig.~\ref{fig:9b}). It
appears at the highest panels because its expansion with height makes it
reach the location of the cut in the MHD cube. We don't aim to reproduce this
neighbouring structure, but only the main flux tube.

In the deeper layers of the photosphere (lower panels of Fig.~\ref{fig:23b})
the relatively thick flux tube splits down its center into two parts. The two
separate parts of the flux tube in the lower photosphere merge while
expanding with height. It is interesting that such groups of flux
concentrations tend to behave like a single flux tube higher up in the
atmosphere owing to expansion and the decrease of $\beta$ with height.

The splitting of the flux tube in the lowest panel leads to a decrease of the
horizontally averaged field strength at this height compared to the
second-lowest panel. It is seen in the framework of the thin flux tube model
as an expansion of the flux tube with depth and produces positive values of
$h_{2}$ (Eq.~\ref{eqn:x31}), clearly seen in the lowest panel.

\begin{figure}
\includegraphics[width=0.95\linewidth,bb= 70 92 520 610]{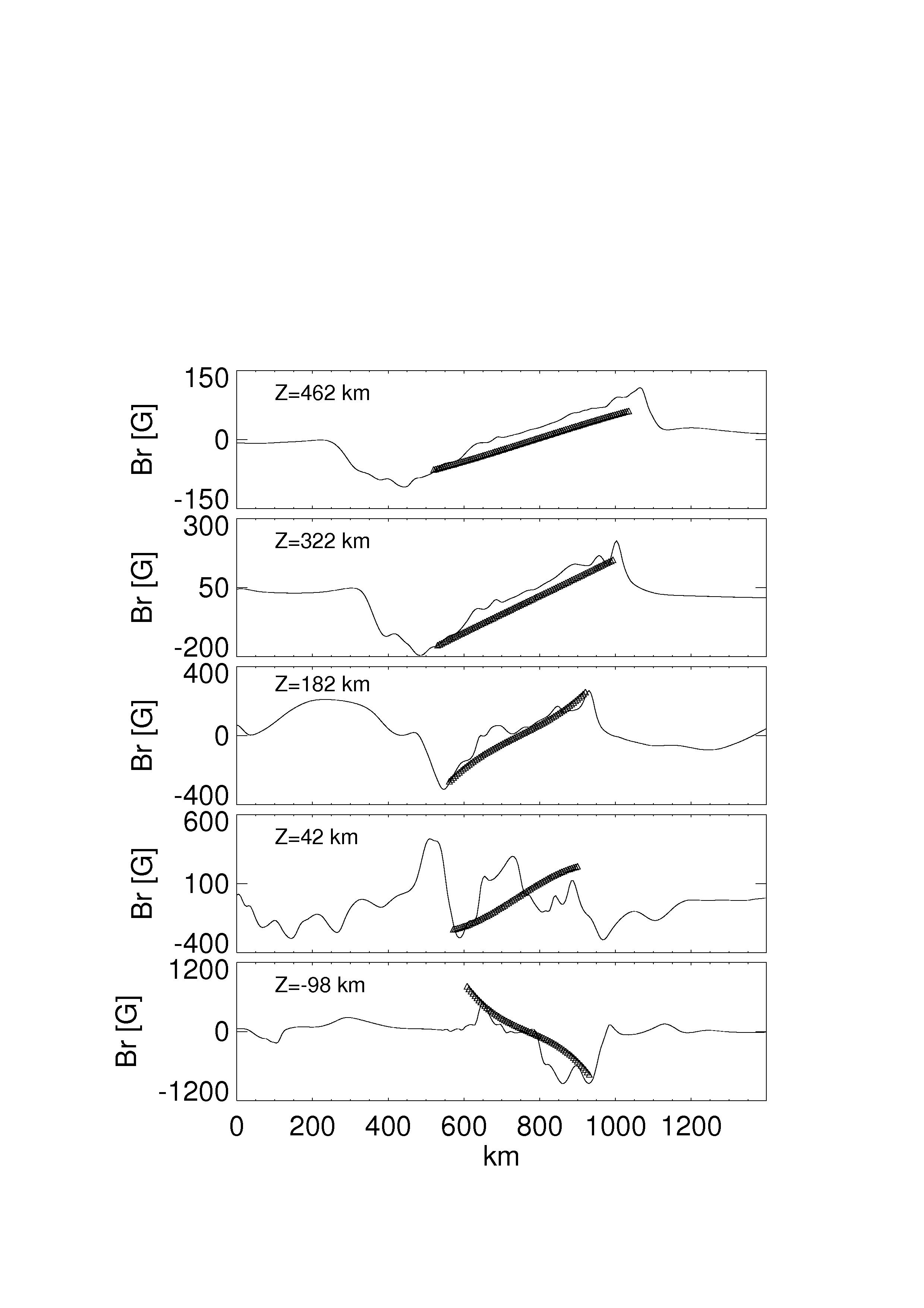}
\caption{Radial component of the magnetic field across the dark line
in the lower right corner of Fig.~\ref{fig:12}. Solid lines represent $B_{r}$ from the simulation, and
triangles are from the thin flux tube model.\label{fig:24b}}
\end{figure}

The radial component of the magnetic field fits reasonably well with the thin
flux tube model for the three upper panels (Fig.~\ref{fig:24b}), except at
the left edge where the small magnetic feature has merged with the main flux
tube. In the two lower panels the actual profiles of $B_r$ are disturbed by
the double structure of the flux tube. In this case the thin flux tube model
cannot be expected to reproduce the actual profiles. In the lowest panel,
$B_r$ from the thin flux tube model has a negative slope due to the expansion
of the flux tube with depth, which leads to negative values of $f_{1}$
(Eq.~\ref{eqn:x32}).

The $4^{th}$-order terms remain very small at all altitudes compared to lower
orders (Table~\ref{table:3}). At the three upper altitudes, the
$3^{rd}$-order contribution is clearly less marked than the $1^{st}$-order
one. The uneven flux distribution at the two lower altitudes results in
somewhat higher contributions of the $2^{nd}$-, $3^{rd}$- and $4^{th}$-
orders compared to the situation at higher altitudes.

\subsection{Very thin flux concentration} \label{sec:7}

The thin flux tube/sheet model is generally thought to be best suited to
describe the smallest flux concentrations in the MHD simulations. This
picture is appropriate for the ideal case where flux tubes/sheets have an
extremely thin boundary layer (separating magnetic and non-magnetic regions)
and for a static plasma. The situation in the photosphere is clearly
different. There vigorous convective flows induce considerable distortions of
very thin flux concentrations. As a consequence, the shape and flux density
distribution of the thinnest magnetic elements may differ significantly from
a thin flux tube/sheet model.

In order for a flux concentration to evolve as a coherent structure in a
plasma with density $\rho$ and velocity $V$, its magnetic energy density
($B^2 / (8 \pi)$) has to be larger than the kinetic energy density of the
flow ($0.5 \rho V^2$). In other words, the magnetic field has to be such that
$B> B_{eq} = V \sqrt{4 \pi \rho}$, where $B_{eq}$ is the equipartition field
strength.

At the surface of the sun we have $B_{eq} \simeq 500$ G. This is a limit
below which we cannot expect to obtain a structure coherent enough to be
described by the thin flux tube/sheet model. Thus we only consider thin
magnetic features with $B> B_{eq} > 500$ G (see contours on
Fig.~\ref{fig:9b}). We also require that flux concentrations remain coherent
at higher altitude (see top-left panel in Fig.~\ref{fig:12}) and are not
located in a region close to opposite-polarity fields, since at these
locations the field morphology gets complicated.

\begin{figure}
\includegraphics[width=0.95\linewidth,bb= 85 115 500 610]{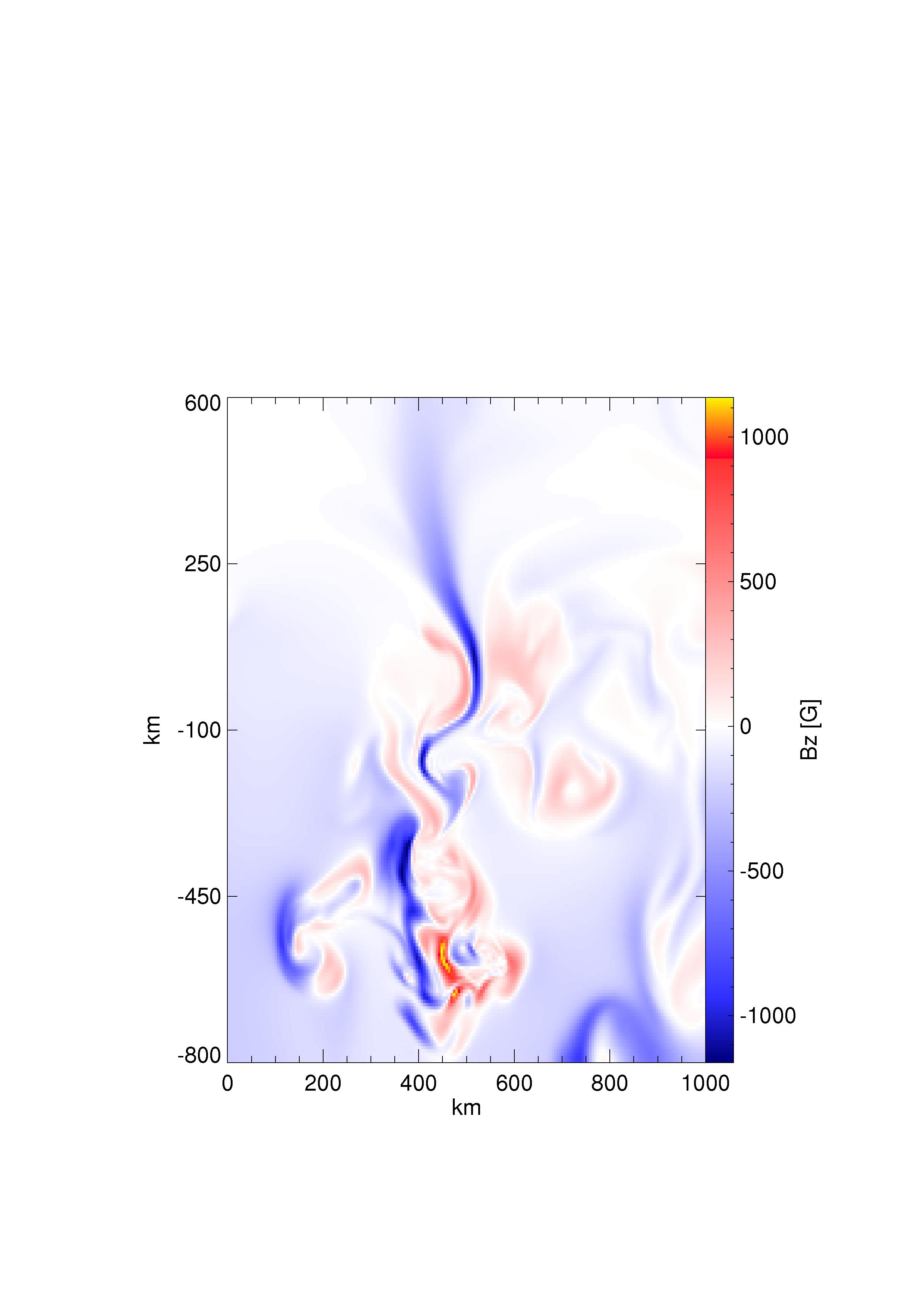}
\caption{Vertical 2D cut in the MHD data showing $B_z$ at the
location indicated by the black line in the upper right part of the maps in Fig.~\ref{fig:9b}. \label{fig:31b}}
\end{figure}

The $B_{z}$ map in the upper left panel of Fig.~\ref{fig:12} indicates that
only relatively few very-thin flux concentrations (that have not merged with
larger magnetic features) are noticeable at the greatest height. We select
one of them, at the location shown by the black vertical line in the
upper-right part of the maps in Fig.~\ref{fig:9b}.

A lateral 2D view of this thin structure (Fig.~\ref{fig:31b}) shows that it
is asymmetric and distorted, with an inclination that varies strongly with
height. Since the magnetic energy density is not far above the equipartition
value, the convective flows influence the morphology of the thin flux
concentration rather strongly. This does not favour of the representation of
very thin flux concentrations in terms of thin flux tube/sheet models.

\section{Conclusions} \label{sec:8}

The total pressure diagnostic (Sect.~\ref{sec:5}) indicates that $P_{tot}$ is
nearly constant across most flux concentrations near the solar surface. This
is a necessary condition for applying the $0^{th}$-order thin flux tube/sheet
approximation. In the higher parts of the atmosphere, tension forces become
important due to the curved field lines and low plasma $\beta$. In this case,
higher orders in the thin flux tube/sheet model are needed to describe flux
elements.

For a detailed analysis of magnetic features in the MHD simulation, we have
adopted two models (thin flux tube and thin flux sheet) depending on the
geometry of the studied flux concentration. We have seen that for flux
concentrations with magnetic field well above the equipartition distribution
(Sects~\ref{sec:4} and~\ref{sec:6}), the models reproduce reasonably well
$B_{z}$ and $B_{x}$ (or $B_{r}$) of the simulated flux concentrations. This
was especially the case in the higher part of the atmosphere. The fits were
less good in the lower part of the atmosphere due to higher $\beta$ and the
vigorous convective flows. In this case, it is rather the overall shape of
$B_{z}$ that is consistent with the approximation. The $2^{nd}$-order terms
of the thin flux tube/sheet approximation contribute at the $5$-$15$ percent
level especially in the upper part of the atmosphere. The $3^{rd}$-order
terms provide a relatively small contribution to $B_{x}$ or $B_{r}$, while
the $4^{nd}$-order terms give a very small contribution to $B_{z}$. This
justifies neglecting the $4^{th}$-order terms and the view that higher-orders
contribute less and less to $B_{z}$ and $B_{x}$ (or $B_{r}$).

In the case of very thin flux concentrations which generally have energy
densities lower than or at most somewhat higher than the equipartition value,
field lines are distorted and partly driven by plasma motions. This leads to
distorted or incoherent flux concentrations which do not have the necessary
symmetry and regularity to be reproduced by a thin flux tube/sheet model. To
what extent these low field strengths are due to the limited resolution (low
Reynold's number) of the simulations still needs to be established. Note,
however, that it has been pointed out \citep{Venkatakrishnan:1986} that the
convective collapse mechanism, thought to be responsible for the
concentration of magnetic flux to kG strengh
\citep{Parker:1978,Spruit:1979,Grossmann:1998}, becomes less efficient as the
amount of magnetic flux per feature decreases. A decrease in field strength
with decreasing magnetic flux has been observationally confirmed
\citep{Solanki:1996}.

\textbf{Acknowledgements:} This work was partly supported by the WCU grant
No. R31-10016 from the Korean Ministry of Education, Science and Technology.
L.Y.C. is thankful to the Max-Planck-Institut für Sonnensystemforschung,
Katlenburg-Lindau for a stipend of the International Max Planck Research
School on Physical Processes in the Solar System and Beyond.

\bibliographystyle{aa}
\bibliography{yelles}

\end{document}